\documentclass{article}

\usepackage{microtype}
\usepackage{graphicx}
\usepackage{subfigure}
\usepackage{booktabs} 
\usepackage{hyperref}
\usepackage{balance} 
\usepackage{listings}
\usepackage{xcolor}

\usepackage[accepted]{conference2025}

\usepackage{amsmath}
\usepackage{amssymb}
\usepackage{mathtools}
\usepackage{amsthm}

\usepackage[capitalize,noabbrev]{cleveref}

\theoremstyle{plain}

\theoremstyle{definition}

\theoremstyle{remark}

\usepackage[textsize=tiny]{todonotes}

\conferencetitlerunning{LLM-Guided Credit Assignment in Multi-Agent Reinforcement Learning}

\begin{document}

\twocolumn[
\conferencetitle{Speaking the Language of Teamwork: LLM-Guided Credit Assignment in Multi-Agent Reinforcement Learning}

\begin{conferenceauthorlist}
\conferenceauthor{Muhan Lin}{yyy}
\conferenceauthor{Shuyang Shi}{yyy}
\conferenceauthor{Yue Guo}{yyy}
\conferenceauthor{Vaishnav Tadiparthi}{comp}
\conferenceauthor{Behdad Chalaki}{comp}
\conferenceauthor{Ehsan Moradi Pari}{comp}
\conferenceauthor{Simon Stepputtis}{yyy}
\conferenceauthor{Woojun Kim}{yyy}
\conferenceauthor{Joseph Campbell}{sch}
\conferenceauthor{Katia Sycara}{yyy}
\end{conferenceauthorlist}

\conferenceaffiliation{yyy}{School of Computer Science, Carnegie Mellon University, Pittsburgh, USA}
\conferenceaffiliation{comp}{Honda Research Institute, Ann Arbor, USA}
\conferenceaffiliation{sch}{Department of Computer Science, Purdue University, West Lafayette, USA}

\conferencecorrespondingauthor{Muhan Lin}{muhanlin@cs.cmu.edu}

\vskip 0.3in
]

\printAffiliationsAndNotice{}  

\begin{abstract}
Credit assignment, the process of attributing credit or blame to individual agents for their contributions to a team’s success or failure, remains a fundamental challenge in multi-agent reinforcement learning (MARL), particularly in environments with sparse rewards. Commonly-used approaches such as value decomposition often lead to suboptimal policies in these settings, and designing dense reward functions that align with human intuition can be complex and labor-intensive. In this work, we propose a novel framework where a large language model (LLM) generates dense, agent-specific rewards based on a natural language description of the task and the overall team goal. By learning a potential-based reward function over multiple queries, our method reduces the impact of ranking errors while allowing the LLM to evaluate each agent's contribution to the overall task. Through extensive experiments, we demonstrate that our approach achieves faster convergence and higher policy returns compared to state-of-the-art MARL baselines.
\end{abstract}

\section{Introduction}
Multi-agent reinforcement learning (MARL) has gained significant attention for its ability to model and solve complex problems involving multiple interacting agents. From coordinating autonomous vehicles~\cite{shalev2016safe, zhang2024multi} in traffic systems~\cite{wiering2000multi, chu2019multi} to managing resources in distributed networks, MARL provides a framework for agents to learn optimal policies through interaction with the environment and each other. 
It is common practice in MARL to learn decentralized policies which operate over local observations, so as to avoid exponential scaling in the joint state-action 
\begin{figure}[t]
\vskip 0.2in
\begin{center}
\centerline{\includegraphics[width=\columnwidth]{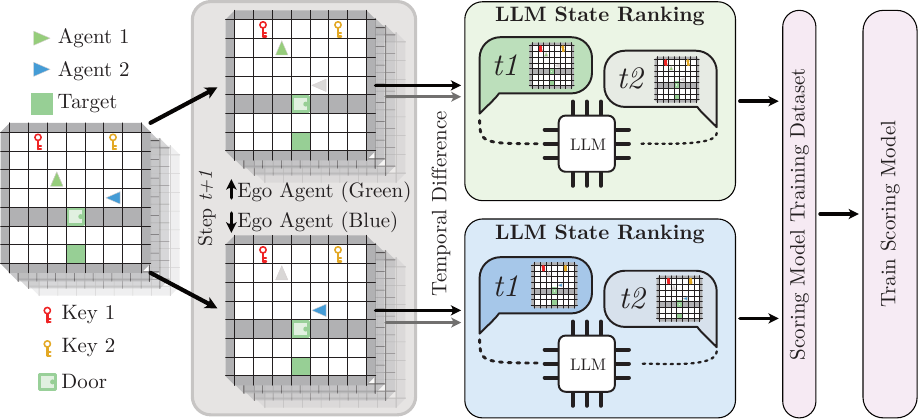}}
\caption{Overview of our method LCA:
    We first generate the agent-specific encodings of state observations, and then prompt an LLM to execute pairwise state ranking from each agent's perspective in the contexts of collaboration. Specifically, if ranking state pairs in Agent 1's perspective, Agent 1 will be encoded as the ``ego" agent and other agents as ``teammates" in the observation, allowing the LLM to differentiate them with the language-based observation description. The individual rewards trained with such agent-specific ranking results properly handle the credit assignment in MARL. We test our approach in the grid world and pistonball environments. }
\label{fig:envs}
\end{center}
\vspace{-3ex}
\vskip -0.2in
\end{figure}
space of all the agents. Decentralization can lead to training instability, however, since the environment appears to be non-stationary from the perspective of each agent as each agent's policy is changing over time.

The centralized-training-decentralized-execution (CTDE) paradigm~\cite{lowe2017multi, kim2023variational, kim2023adaptive} effectively solves this problem by leveraging global state and joint action information during training.
However, one of the fundamental challenges in MARL is the credit assignment problem~\cite{foerster2018counterfactual, rashid2020monotonicvaluefunctionfactorisation}: determining how to attribute the team's success or failure to individual agents' actions.
In single-agent reinforcement learning, the reward signal directly reflects the consequence of the agent's actions, facilitating straightforward learning of optimal policies.
In contrast, MARL involves multiple agents whose actions collectively influence the \textit{team reward}, making it difficult to discern each agent's individual contribution.


Credit assignment can be implicitly addressed through the use of value decomposition methods~\cite{rashid2020monotonicvaluefunctionfactorisation, sunehag2017value, son2019qtran} in CTDE.
These approaches decompose the team value into a (possibly nonlinear) combination of per-agent values. Despite their successes~\cite{wang2020qplex}, such decompositions are less competitive in sparse reward settings where feedback is infrequent and often delayed~\cite{liu2023lazy}.
Such drawback limits the application of these methods, as sparse reward settings remain exceedingly common largely due to the difficulty of crafting dense, value-aligned reward functions~\cite{leike2018scalable, skalse2022defining, knox2023reward, booth2023perils}.

Recent work has shown that large language models (LLMs) can be used to autonomously generate preference rankings and learn dense reward functions~\cite{lee2023rlaif}.
While such techniques have been shown to aid learning in the presence of sparse rewards in single-agent settings~\cite{lin2024navigating}, it remains an open question as to whether AI-generated reward functions can properly attribute credit in the multi-agent case. This work seeks to answer the question: \textbf{can we leverage LLMs to assign credits in MARL by generating informative, agent-specific rewards based on natural language descriptions of tasks and goals?}

In this paper, we propose \textbf{L}LM-guided \textbf{C}redit \textbf{A}ssignment \textbf{(LCA)}, a novel framework that integrates LLMs into the MARL training process to facilitate credit assignment with sparse rewards. Our approach retrieves information about the overall team objective and its key steps from existing team rewards and provides them to the LLM.
The LLM generates preference rankings over each agent's actions so that actions that are more contributive from the perspective of the team's success are preferred. These rankings are used to train dense potential-based reward functions for each agent, simultaneously addressing both credit assignment and reward sparsity.

We conduct extensive experiments in various MARL environments characterized by sparse rewards and complex agent interactions.
Our results show that agents trained with our LLM-generated, agent-specific rewards achieve faster convergence to optimal policies and higher overall returns compared to agents trained with hand-crafted dense agent-specific rewards.
Furthermore, we demonstrate that our framework is resilient to ranking errors, allowing for the effective use of smaller, more accessible language models without significant performance degradation.
Our work makes the following key contributions:
\begin{enumerate}
    \item We leverage LLMs to generate dense agent-specific rewards based on a natural language description of the team's goal, successfully handling the credit assignment.

    \item We empirically show that our approach leads to higher policy returns and faster convergence speeds than baseline methods, even when rankings are generated from smaller, more error-prone LLMs.

\end{enumerate}

\section{Related Works}
Credit assignment in multi-agent reinforcement learning remains a fundamental challenge, especially in environments with sparse team rewards. There are two main classes of traditional approaches for the credit assignment, value decomposition \cite{sunehag2017value, rashid2020monotonicvaluefunctionfactorisation, du2019liir, foerster2018counterfactual} and slight modifications to known algorithms such as the gradient-descent decomposition \cite{su2020valuedecompositionmultiagentactorcritics}. 

There are also some works that combine the basic ideas of both method classes. \cite{kapoor2024assigningcreditpartialreward} adapts the partial reward decoupling into MAPPO \cite{yu2022surprisingeffectivenessppocooperative} to eliminate contributions from other irrelevant agents based on attention. The other work \cite{wen2022multiagentreinforcementlearningsequence} utilizes transformer with PPO loss \cite{schulman2017proximal} adapted to the value decomposition idea.

The methods above have made progress in assigning credits, but their effectiveness diminishes with delayed or sparse rewards. For example, the work \cite{liu2023lazy} shows the poor performance of QMIX \cite{rashid2020monotonicvaluefunctionfactorisation} with sparse rewards. However, designing dense rewards to combat this challenge is difficult given the complexity of tasks ~\cite{leike2018scalable, knox2023reward, booth2023perils}. Although social-influence-based rewarding calculates dense individual rewards \cite{jaques2019social}, it requires teammates' behavior models, which often need additional training to estimate and update.

One general method of generating dense rewards, particularly in single-agent settings, is Reinforcement Learning with Human Feedback (RLHF) \cite{christiano2017deep} and its extension, Reinforcement Learning with AI Feedback (RLAIF) \cite{leerlaif}. These methods have been successfully applied in domains like text summarization and dialogue generation \cite{ziegler2020finetuninglanguagemodelshuman}, where human or AI-generated feedback is used in training in the absence of clear environmental rewards. However, these approaches are limited to single-agent environments and do not address the unique requirements and challenges that exist within the multi-agent counterparts, according to RLAIF. \cite{zhangsimple} shows one direction of generating dense rewards for credit assignment with LLM in multi-agent scenarios. Utilizing the coding capabilities of LLM, this method iteratively queries LLM to generate multiple reward functions with high density and refine the reward functions gradually in the training process. However, this method can suffer from LLM hallucination problems, which can cause misleading rewards due to inconsistent rankings or other ranking errors. Considering these problems, our method adapts the potential-based RLAIF \cite{lin2024navigating}, which can handle LLM hallucination with the multi-query approach, from the single-agent scenarios to multi-agent ones, and successfully handles the credit assignment problem.

\section{Background}

\textbf{Multi-Agent Reinforcement Learning:} 
~~We consider a fully cooperative Markov Game~\cite{matignon2012independent}, which generalizes the Markov Decision Process (MDP) to multi-agent settings where multiple agents interact in a shared space and collaborate by maximizing a common reward function. A fully cooperative Markov Game is represented by the tuple $(N, S, \{A_i\}_{i=1}^N, P, R, \gamma)$, where $N$ is the number of agents, $S$ represents the set of global states, $\{A_i\}$ is the action space for each agent, and $P(s'|s, a_1, \dots, a_N)$ describes the probability of transitioning from one state to another based on the joint actions of all agents. The agents share a reward function $R(s, a_1, a_2, \dots, a_N)$, which assigns a common reward based on the state-action pairs. The objective is for the agents to collaboratively learn policies that maximize the cumulative discounted reward, where $\gamma$ denotes the discount factor.

\noindent\textbf{Value Decomposition:} In the context of multi-agent systems, value decomposition allows each agent to independently learn a value function, with all value functions collectively working toward a common goal or outcome. Value decomposition refers to the process of decomposing a complex global value function into multiple components. Each local component can then be optimized independently, while still contributing to the global target.

\noindent\textbf{Preference-Based Reinforcement Learning:} The underlying framework of our work is preference-based reinforcement learning, where preference labels over agent behaviors are used to train reward functions for RL policy training \citep{christiano2017deep, ibarz2018reward, lee2021pebble, lee2021b}. Given a pair of states $(s_a, s_b)$, an annotator labels preference $y \in \{0,1\}$ to indicate which state is closer to the task goal: $y=0$ if $s_a$ is ranked higher than $s_b$, and $y=1$ if $s_b$ is ranked higher than $s_a$.
 
We introduce a parameterized state-scoring function $\sigma_\psi$, often referred to as the potential function and typically identified with the reward model $r_\psi$. Based on this, the probability that the $s_a$ is ranked higher than $s_b$ is computed with the standard Bradley-Terry model \citep{bradley1952rank}, \begin{equation} \begin{aligned} P_\psi[s_a \succ s_b] &= \frac{\exp \left(\sigma_\psi(s_a) \right)}{\exp \left(\sigma_\psi(s_a) \right) + \exp \left(\sigma_\psi(s_b) \right)}\ \\ &= \text{sigmoid}(\sigma_\psi(s_a) - \sigma_\psi(s_b)), \end{aligned} \end{equation} Utilizing a preference dataset $\mathcal{D} = \{(s_a, s_b, y) | s_a, s_b \in \mathcal{S}\}$, preference-based RL trains the state-scoring model $\sigma_\psi$ via minimizing the cross-entropy loss. This process aims to maximize the score difference between higher-ranked and lower-ranked states: 
\begin{equation} 
\begin{aligned} 
\mathcal{L} &= -\mathbb{E}_{(s_a, s_b, y) \sim \mathcal{D}} \bigg[ \mathbb{I}\{y = (s_a \succ s_b)\} \log P_\psi[s_a \succ s_b] \\
& + \mathbb{I}\{y = (s_b \succ s_a)\} \log P_\psi[s_b \succ s_a]\bigg], 
\end{aligned} 
\label{eq} 
\end{equation} 
with $\mathbb{I}{\cdot}$ as the indicator function. This framework is applied in both Reinforcement Learning from Human Feedback (RLHF) and Reinforcement Learning from AI Feedback (RLAIF), where the outputs of the state-scoring model are directly used as rewards. The primary difference between these approaches lies in the choice of annotator—either a human or a large language model (LLM). 

Using LLMs for preference labeling reduces human labor but with inevitable ranking errors, resulting in misleading rewards and inefficient training. One critical source of errors is inconsistent rankings on the same state pairs across multiple prompting trials when the LLM is uncertain about their preference. It is proven that formulating potential-based RLAIF rewards as $r(s_t,s_{t+1}) = \sigma_\psi(s_{t+1}) - \sigma_\psi(s_t)$, instead of $\sigma_\psi(s_t)$, causes $r(s_t, s_{t+1})$ to converge to 0 as LLM uncertainty increases \cite{lin2024navigating}. Such uninformative reward effectively mitigates the negative impact of inconsistent rankings.

\section{Method}

Existing RLAIF approaches~\cite{lin2024navigating} do not lend themselves well to multi-agent settings when ranking joint state-actions.
Consider a two-agent scenario in which the agents perform actions with conflicting contributions toward the team goal: one positive and one negative.
The positive reward from a beneficial action that contributes to the team's success is canceled out by the negative reward from another agent.
This results in an ambiguous state which is difficult for an LLM to rank when considering both agents, ultimately resulting in a sparse rather than dense reward function.
In contrast, our LCA approach seeks to decompose the joint preference ranking into individual preference rankings for the purpose of learning individual reward functions, overcoming this issue.

\subsection{LLM-based Reward Decomposition}
\noindent\textbf{\textit{Describing Team Goals from Team Rewards:}} 
Given that not all environments provide explicit, natural language descriptions of states, goals, or sub-goals, this information can be inferred from the team reward structure by investigating a trajectory sampled beforehand.
Without loss of generalization, we assume that there exists one team reward function, $r_t(s_i)$, from the environment, which is usually sparse (We assume it does not include step penalty and is not finely hand-crafted).
Therefore, on a trajectory randomly sampled without a limit of max steps - which means it ends when the team task is completed - there are only a few states $s_i$ where $r_t(s_i) \neq 0$. If $r_t(s_i) > 0$, $s_i$ should be a key landmark of completing the team task.
If $r_t(s_i) < 0$, it would be critical to avoid this state $s_i$. Therefore, the natural language description of such $s_i$ following the order they appear on the sampled trajectory can provide LLM enough information about how the agent team should complete the task, which will be critical information for agent-specific state ranking.\\

\noindent\textbf{\textit{Agent-specific State Ranking:}} We prompt an LLM to implicitly assign credits to each agent separately by ranking state pairs based on the agent's own actions from that agent's perspective. We first generate an agent-specific encoding $o^i$ of the observation $o$ of a state $s$ by labeling the agent $i$ itself as the ``ego" and any other agent as the ``teammate", allowing the LLM and state-scoring models to identify which agent they are evaluating. Given any state transition $(s, a, s')$, where $a = \langle a_1, \dots, a_n \rangle$ and $n$ is the number of agents, the LLM generates a preference label for agent $i$ as:
\[
y^i(s, a, s') = y^i(o^i, a_i, o'^i).
\]
The LLM is then prompted to reason from agent $i$'s perspective to determine whether the agent’s action $a_i$ between these two states, $o^i$ and $o'^i$, is appropriate for collaboration. If agent $i$ performs a correct action while another agent $j$ performs an incorrect one—a scenario where single-agent-style RLHF struggles to generate a single ranking—this method assigns:
\[y^i(s, a, s') = (o'^i \succ o^i) = (s' \succ s),\] and \[ y^j(s, a_j, s') = (o^j \succ o'^j) = (s \succ s').\]
\noindent\textbf{\textit{LLM-Guided Individual Reward Training:}}
Given that the LLM implicitly assigns credit by generating differentiated rankings for each agent $i$ $\mathcal{D}^i = \{(s_a, s_b, y^i) | s_a, s_b \in \mathcal{S}\}$, these rankings can be used to train individual state-scoring models $\sigma^i(o^i)$. The loss function for each individual state-scoring model will be
\begin{equation} 
\begin{aligned} 
\mathcal{L}^i &= -\mathbb{E}_{(s_a, s_b, y^i) \sim \mathcal{D}^i} \bigg[ \mathbb{I}\{y^i = (s_a \succ s_b)\} \log P_\psi ^i[o_a^i \succ o_b^i] 
\\& + \mathbb{I}\{y^i = (s_b \succ s_a)\} \log P_\psi ^i[o_b^i \succ o_a^i]\bigg], \\
&= -\mathbb{E}_{(s_a, s_b, y^i) \sim \mathcal{D}^i} \bigg[ conf\{y^i = (s_a \succ s_b)\} \\ & \log (sigmoid(\sigma^i_\psi(o_a^i) - \sigma^i_\psi(o_b^i))) + \\& 
conf\{y^i = (s_b \succ s_a)\} \log (sigmoid(\sigma^i_\psi(o_b^i) - \sigma^i_\psi(o_a^i))) \bigg].
\label{push0}
\end{aligned} 
\end{equation} 
The individual reward will be formulated as
\begin{equation} 
r_i(s,a_i,s') = \sigma_\psi ^i(o'^i) - \sigma_\psi ^i(o^i)
\end{equation} 
except the case where the agent $i$ stays still without taking an actual action and the reward will be 0. 

This reward function generalizes potential-based rewards from single-agent to multi-agent settings, while maintaining the claims in~\cite{lin2024navigating} that the RLAIF loss encodes ranking confidence, and that inconsistent rankings, implying that the confidence of two possible ranking results over a state pair are closer, possible drive the individual reward towards zero with the loss function of the state-scoring model in Eq. ~\ref{push0}. Intuitively, this means that the individual reward functions are robust to ranking errors stemming from high uncertainty when each state-action pair is ranked multiple times.

Additionally, it is unnecessary to train one reward function for each agent if agents are homogeneous with the same individual task. Since these agents take the same, exchangeable role in the team, for a transition $(s_a, a, s_b)$ with encoded observation $o_a^i, o_b^i$ for agent i, there must exist another transition $(s_a', a, s_b')$ with encoded observation $o_a'^j, o_b'^j$ for agent $j$ such that $o_a^i = o_a'^j, o_b^i = o_b'^j$. The loss function for agent $i$'s state-scoring model over the preference dataset $\mathcal{D}^i$ can be written as
\begin{equation} 
\begin{aligned} 
\mathcal{L}^i &= -\mathbb{E}_{(s_a, s_b, y^i) \sim \mathcal{D}^i} \bigg[ \mathbb{I}\{y^i = (o_a^i \succ o_b^i)\} \log P_\psi ^i[o_a^i \succ o_b^i] 
\\& + \mathbb{I}\{y^i = (o_b^i \succ o_a^i)\} \log P_\psi ^i[o_b^i \succ o_a^i]\bigg], \\
&= -\mathbb{E}_{(s_a', s_b', y^i) \sim \mathcal{D}^i} \bigg[ \mathbb{I}\{y^i = (o_a'^j \succ o_b'^j)\} \log P_\psi ^i[o_a'^j \succ o_b'^j] 
\\& + \mathbb{I}\{y^i = (o_b'^j \succ o_a'^j)\} \log P_\psi ^i[o_b'^j \succ o_a'^j]\bigg].
\end{aligned}
\end{equation} 
If agent $i$ and $j$ share $y^i, \mathcal{D}^i, P_\psi^i$, which means they share the ranking dataset and the state-scoring model,  $\mathcal{L}^i$ will be directly transformed to $\mathcal{L}^j$. Therefore, homogeneous agents with the same tasks can be grouped together and share the same reward function. The single reward function can handle the credit assignment among them and gives distinct individual rewards by taking differentiated observations in their own view over the current state. We only need to train different reward functions for heterogeneous agents or homogeneous ones with different pre-assigned tasks.

\subsection{Prompt Designs for Agent-specific State Ranking Reflecting Collaboration}

Although we decompose the joint state-action rankings into individual rankings, it does not mean the ranking for each agent is the same as it would be in a single-agent scenario.
Although the LLM thinks in the ``ego" agent's view, it needs to think for the team rather than the ``ego" agent itself so that the agent-specific ranking can evaluate the collaboration between the ``ego" agent and the ``teammate" agents and correctly assign credit for collaboration. This section introduces how to achieve this via prompt design.

During collaboration, each agent's policy depends on the states and actions of other agents.
We design our prompt to make this dependency understandable by LLMs. We consider two types of collaboration dependencies:
\begin{enumerate}
    \item \textbf{Behavior dependence:} Teammates’ current state and latest action influence the ego's current action choice. 
    \item \textbf{Task dependence:} The ``ego" agent needs to change its task steps according to others’ task requirements.
\end{enumerate}

The Two-Switch and Victim-Rubble environments introduced in the experiment section ~\ref{sec:exp_setup} are two examples corresponding to these collaboration dependencies.
We introduce prompt designs for the above two dependency types separately with these two examples.

\subsubsection{\textbf{\textit{Prompt Design for Behavior Dependence}}}
In the Two-Switch environment (see Sec.~\ref{sec:exp_setup} for description), the optimal teamwork requires two agents to separately trigger each switch and unlock the door. Without specific guidance and inter-agent communication, it is natural that the ``ego" agent will observe which switch its teammate is moving towards and then choose the other switch. However, if the teammate is undecided and fails to commit to a particular switch, this can lead to a deadlock as each agent adapts its goal based on the other agent's goal inferred by the teammate's latest action. Such behavior is undesirable as it can introduce non-stationarity into the environment, i.e. from the perspective of the ``ego" agent the teammate's behavior can rapidly change as its policy updates.
In addition to destabilizing training, this kind of behavior dependency can create sub-optimal policies in which one agent is ``lazy'' and fails to contribute to the team's success~\cite{liu2023lazy}.

The agent-specific LLM-generated rankings produced by our approach are designed to address these issues. We instruct the LLM to \textbf{believe the teammates are acting with the optimal policy} when generating rankings. The resulting individual reward function will encourage the ``ego" agent to pursue optimal actions under the assumption that the teammates will act similarly optimal.
In this way, the agents avoid falling into both deadlocks and behaviors where they must compensate for ``lazy'' teammate behavior. To achieve this, besides offering the team target, key steps, the environment, and current states of all agents, we add the following constraint to our prompt:\\

\noindent\textit{\textbf{Assuming the ``teammate" agent will take the best action for the team at this step,} does the current action taken by the ``ego" agent help ... from the view of team?}\\

With this prompt, the LLM understands that it should rank state pairs based on whether the ``ego" has made the optimal decision, without being influenced by or hesitating over the teammate’s subsequent actions. 

\subsubsection{\textbf{\textit{Prompt Design for Task Dependence}}} In the Victim-Rubble environment (see Sec.~\ref{sec:exp_setup} for description), optimal teamwork requires two agents to adjust the order of their task steps in response to the needs of their teammate.
Specifically, the green agent must prioritize which victims to heal and the orange agent must prioritize which pieces of rubble to remove.
For example, if the agents start in the center room, then the orange agent should prioritize removing the rubble in the right rooms as it blocks access to a victim which the green agent will need to heal.
And depending on the relative location of the orange and green agents, the green agent may be more optimal by first healing the accessible victim in the left room while it waits for the orange agent to open up access to the blocked victim.

To achieve this level of collaboration, besides offering the description of the team target, key steps, the environment and the current states of agents, the prompt should first describe the dependency between different agents’ tasks:
\begin{lstlisting}[basicstyle=\scriptsize\ttfamily]
The green agent always prioritizes rescuing victims whose path is free of any rubble, waiting for the orange agent to remove rubbles and clear paths. 
\end{lstlisting}
Then describe the current dependency constraints:
\begin{lstlisting}[basicstyle=\scriptsize\ttfamily]
Rubble1: Chamber5 (8,1), **blocking the only passage** between Chamber5 and 3 from the side of Chamber5
Rubble2: Chamber5 (9,2), **blocking the only passage** reaching Chamber4 which contains one Victim
\end{lstlisting}
And also tell LLM which role the ``ego" agent takes:
\begin{lstlisting}[basicstyle=\scriptsize\ttfamily]
You are the orange agent at Chamber3 (4,3)
\end{lstlisting}
Combining these information, the LLM can identify the next rubble the orange agent should first remove. Part of the example response is as follows:
\begin{lstlisting}[basicstyle=\scriptsize\ttfamily]
The next step for the orange agent should be to clear the path to Chamber4 so that the green agent can rescue the victim.
\end{lstlisting}
\section{Experiments}
We tested LCA in three multi-agent collaboration scenarios without inter-agent communication, outer access to policy models, or state transition models. Fig.~\ref{fig:multigrid} shows the layouts.
\subsection{Experiment Setup}
\vspace{-3ex}
\label{sec:exp_setup}
\begin{figure}[ht]
\vskip 0.2in
\begin{center}
\centerline{\includegraphics[width=\columnwidth]{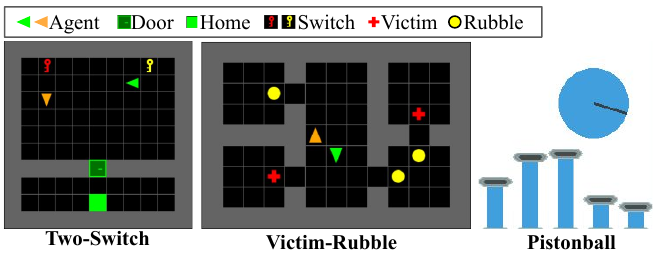}}
\caption{Grid world environments with Two-Switch (left), Victim-Rubble (middle) and Pistonball (right) variants from left to right.}
\label{fig:multigrid}
\end{center}
\vskip -0.2in
\end{figure}

\noindent\textbf{Grid World.} We examine two multi-agent collaboration scenarios within Grid World \cite{swamy2024minimaximalist}: \textbf{Two-Switch} and \textbf{Victim-Rubble}. 

In the Two-Switch variant, two agents (green and orange triangles) start from random positions in the upper room and at least one of them should navigate to the target (green rectangle) in the lower room. 
There is one locked door that blocks the agent's way to the goal and only opens when both switches have been triggered.
To unlock the door, the agents must move to the switches, face the switches and trigger them. Therefore, agents are expected to distribute switches between each other and trigger each switch separately. This should be achieved by observing the other agent's position since agents cannot communicate. 

In the Victim-Rubble variant, the green agent must heal all victims (red crosses) and the orange agent must clear all rubble (yellow circles) in the rooms. There is always one victim lying at the end of a long corridor (the upper-right one in the middle environment in Fig. ~\ref{fig:multigrid}) and there are always two pieces of rubble blocking the way to this victim. Additionally, there is always one piece of rubble blocking nothing and one victim to which the passage is free to pass through. To complete the task as fast as possible, the orange agent should learn to first clear the rubble blocking the passage, and the green agent should learn to first heal the accessible victim.

\noindent\textbf{Pistonball.} We also investigate the Pistonball environment from PettingZoo \cite{NEURIPS2021_7ed2d345}.
There are five pistons which are five independent agents in this environment, moving upwards and downwards. They aim to push the ball starting from the rightmost point of the environment to reach the leftmost point with the least steps.

We compare our approach with the following baselines:
\begin{itemize}
    \item {\bf MAPPO with the default team reward} This is the vanilla case of MARL where no explicit credit assignment is done, utilizing the team's overall objective of each environment with human-specified reward functions given to all agents. In grid world variants, each agent receives a default reward of $0$ for failure and $1$ when the team completes the task. Additionally, both agents earn $1$ if any agent either handles a switch, victim, or rubble, or arrives home. A step penalty of $-n/n_{max}$ is applied, where $n$ is the step count and $n_{max}$ is the episode's maximum time steps. In the Pistonball variant, all agents obtain the team reward of $1$ when the ball reaches the leftmost point, and a step penalty of -0.1 for each step.

    \item {\bf MAPPO with the default team reward plus individual rewards} Besides the team reward based on outcomes (success/failure), this baseline assigns credits in a naive way with default hand-crafted individual rewards. In the Two-Switch variant, the default individual reward is defined as $1$ if the agent triggers a switch or arrives at the goal. In the Victim-Rubble variant, the individual reward is $1$ for the orange agent if it removes a piece of rubble, and for the green agent if it heals one victim. There are no simple individual rewards in the Pistonball variant, which thus does not have this baseline.
    \vspace{-0.5ex}
    \item {\bf QMIX and VDN with the default team reward}
    These two baselines decompose the team reward described above into individual Q values for credit assignment \cite{rashid2020monotonicvaluefunctionfactorisation, sunehag2017value}. We evaluate LCA against these two classical value-decomposition methods to show its effectiveness.
\end{itemize}

Team rewards often fail to discourage poor agent behaviors. While naive hand-crafted individual rewards can partially address this, their sparsity limits effectiveness. Our method's dense individual rewards are expected to significantly outperform these alternatives. Specifically speaking,

1) In Two-Switch: Team rewards grant all agents +1 when a switch is triggered, regardless of which agent triggers it. If the orange agent learns this first, the green agent may remain idle, letting the orange agent trigger both switches and still earning +2. This inefficiency increases team steps. Our rewards immediately penalize agents once they act improperly.

2) In Victim-Rubble: If the orange agent fails to clear the rubble blocking a passage or remains idle, the green agent can only save accessible victims. Both agents still earn +1 team reward for this action, despite reduced overall performance. Our rewards immediately penalize the orange agent once it stops moving toward the critical rubble.

3) In Pistonball: Team rewards penalize all pistons if the ball moves right, even if some act correctly. There are no straightforward individual rewards, unless with extensive tuning. Our dense rewards target only the piston directly responsible for the incorrect ball motion.

These challenging collaborative scenarios make the three environments ideal for testing our method against baseline approaches. Without loss of generality, we employ IPPO as the underlying policy-training framework \citep{schulman2017proximal} and assume the agent has no knowledge of the task before training, i.e., is randomly initialized.

We randomly sampled sequential state pairs to train state-scoring models and formulate potential-difference reward functions in each environment. Since the agents in the Two-Switch environment are homogeneous with the same individual tasks, a single state-scoring model is trained with 4400 state pairs in total for two agents. Similarly, a single state-scoring model is trained with 1000 state pairs in total for five agents in the Pistonball environment. Two state-scoring models are trained for the two heterogeneous agents in the Victim-Rubble variant and each takes 2000 state pairs.

\begin{figure*}[h]
\vskip 0.2in
\begin{center}
\centerline{\includegraphics[width=\textwidth]{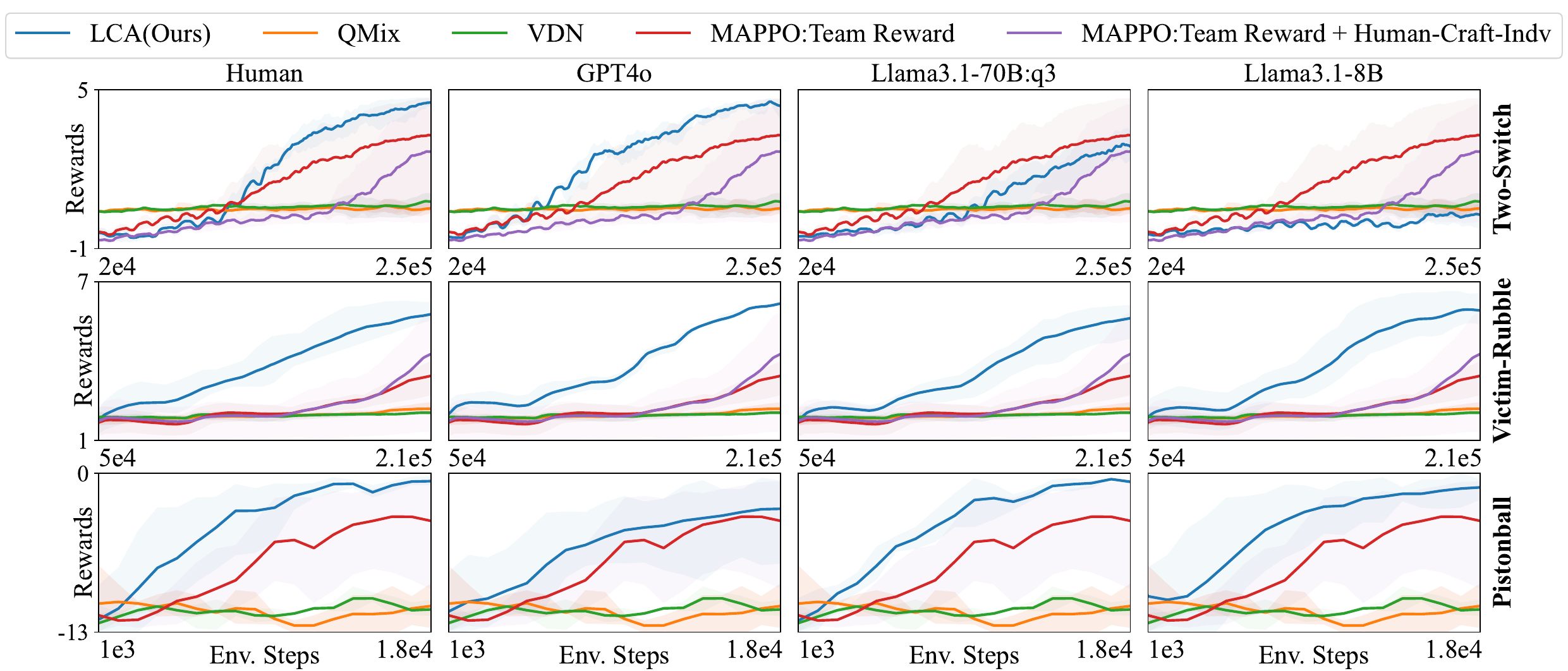}}
\caption{The average learning curves with reward functions trained from single LLM ranking per state pair in the Two-Switch, Victim-Rubble and Pistonball environments over 3 random seeds, with the return variance visualized as shaded areas. The training returns shown as the y axis are measured with vanilla individual rewards plus team rewards.}
\label{fig:1query}
\end{center}
\vspace{-3ex}
\vskip -0.2in
\end{figure*}

\subsection{Single-Query Evaluation}

We first evaluate the performance of our method using a single query to the LLM to rank each sampled state pair. In each environment, we train our state-scoring models with the human ranking-heuristic function, which serves as an estimated ground-truth ranking based on human heuristics, and evaluate them against 3 LLMs: GPT-4~\citep{achiam2023gpt}, and two versions of Llama-3.1~\citep{touvron2023llama}—one small and fast version with 70B parameters, referred to as q3\_K\_M, and another with 8B parameters. Then the potential-difference rewards based on state-scoring models above are employed to train 3 RL policies with random seeds and initializations for each method. The results, as well as the baseline performance, are shown in Fig. ~\ref{fig:1query}.

In the Two-Switch variant, our method with human heuristics and GPT4o achieves the optimal return ($5-step\_penalty$) in 250k training steps with faster learning speed and less variance than baselines. In this single-query experiment, it is normal to observe that policies trained with the quantized Llama3.1-70B:q3 learn more slowly and the rewards from Llama-3.1 8B generating noisy outputs fail to train a useful policy according to \cite{lin2024navigating}. They can be further improved with multiple ranking queries per state pair, particularly Llama 3.1-70B:q3, which outperforms the baselines with just two queries, as demonstrated in the next section on multi-query experiments.

\begin{figure}[ht]
\vskip 0.2in
\begin{center}
\centerline{\includegraphics[width=1\linewidth]{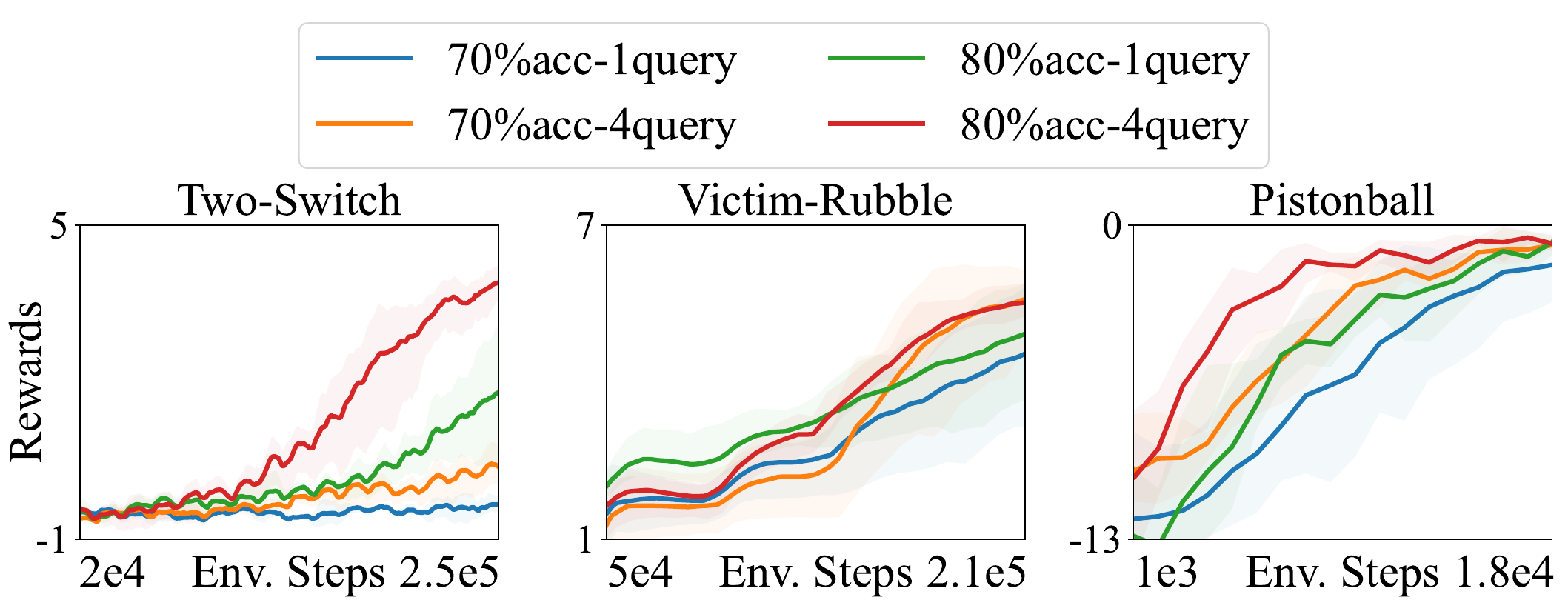}}
\caption{The learning curves with reward functions trained from four-query synthetic experiments over 3 random seeds.}
\label{fig:4query}
\end{center}
\vspace{-3ex}
\vskip -0.2in
\end{figure}

In the Victim-Rubble variant, the default reward easily fails to reach a high return in 210k training steps while LCA with human, GPT4o, Llama3.1-70B:q3 and Llama3.1-8B rankings converges much faster and reaches the optimal reward ($7-step\_penalty$). GPT4o-reward rollout over an episode in Appendix \ref{sec:rollout} shows that LCA effectively decomposes sparse team rewards into dense informative individual rewards. However, the imperfect human ranking heuristic causes our method to learn slightly more slowly than with GPT4o. The human ranking heuristic in this environment forces the green agent to always first save the accessible victim and the orange agent to always first remove the rubbles blocking passages. However, on certain trajectories from suboptimal policies during training, the orange agent may encounter harmless rubble before clearing other rubble, making immediate removal more efficient than returning later. Llama3.1-70B:q3 can have similar ranking flaws. Such ranking flaws may lead to some local optimality and slightly slow down the training speed.

In the Pistonball variant, the baselines fail with the sparse vanilla team reward, while our method with human, Llama3.1-8B and -70B:q3 learn the optimal policy much faster with less variance in 18k training steps. Compared with other LLMs, GPT4o struggles a bit to understand the introduced physical mechanism, so it slows down the training process slightly but still trains some useful policies.

Due to the $\epsilon$-greedy controller PPO methods do not adopt, QMIX and VDN can sometimes start training with a higher return, where $\epsilon=1$, than IPPO-based LCA and MAPPO. However, sparse rewards cause QMIX and VDN to learn slowly and fail to reach significant returns within LCA’s limited training steps, though they learn much faster after a few hundred thousand training steps exceeding LCA training time. We also tried LIIR \cite{du2019liir} and encountered similar consequences, so we ignore its results here.

\subsection{Multi-Query Evaluation}
This section verifies that our method successfully inherits the robustness of potential-based rewards to noisy preference labels, extending the multi-query approach from single-agent scenarios to MARL. The multi-query approach is to query about ranking over each state pair in the ranking dataset multiple times to handle LLM-ranking inconsistencies for small but fast LLMs generating errors.

\begin{figure}[ht]
\begin{center}
\centerline{\includegraphics[width=1\linewidth]{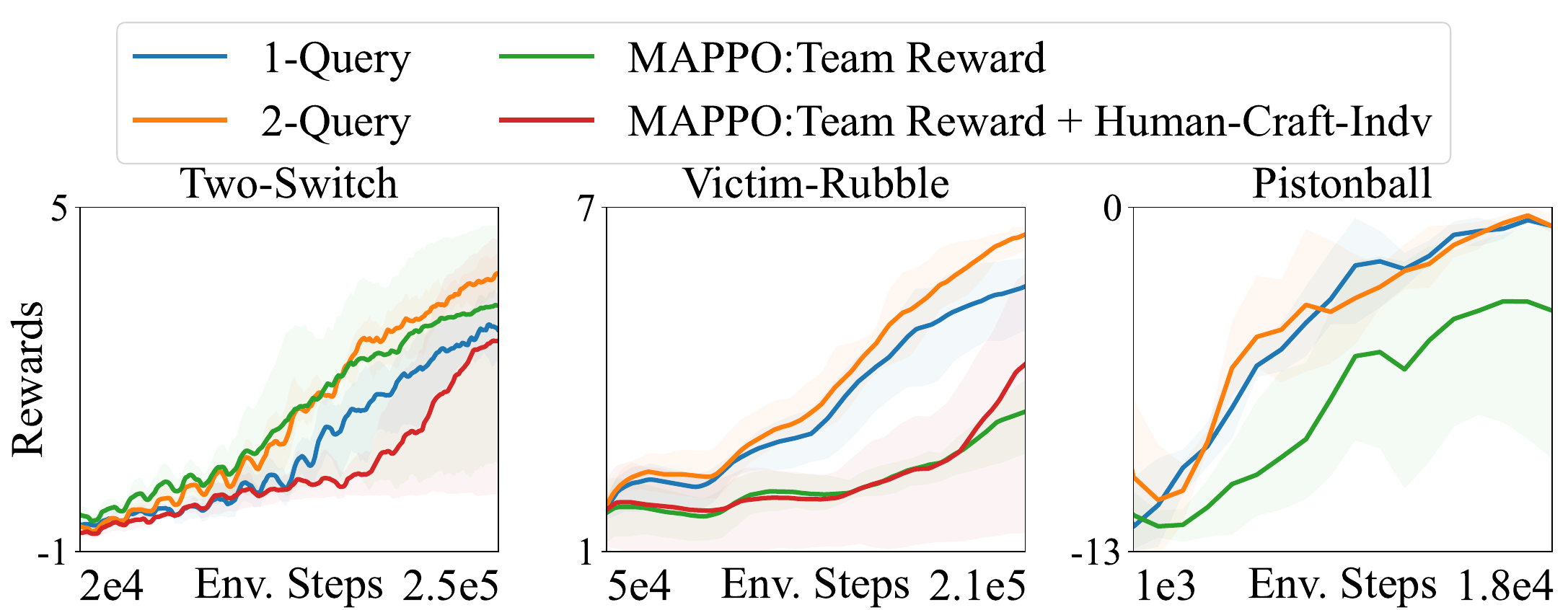}}
\caption{The learning curves with reward functions trained from two-query with Llama3.1-70B:q3 over 3 random seeds.}
\label{fig:2query}
\end{center}
\vspace{-3ex}
\vskip -0.2in
\end{figure}

\subsubsection{\textbf{\textit{Synthetic Ranking Evaluation}}} To evaluate LCA's robustness, we synthesized ranking datasets with 70\% and 80\% accuracy and simulated ranking results with four queries per state pair across three environments. These rankings have correctness between 60\% (near random guessing) and 90\% (high accuracy), providing a comprehensive assessment of LCA’s performance. The four-query ranking datasets are synthesized based on four copies of the human-ranking datasets by randomly flipping a specific percentage of rankings. The data are used to train state-scoring models separately, based on which we obtain multi-query potential-based rewards. Fig. ~\ref{fig:4query} shows the resulting policy learning curves averaged over 3 random seeds. We can see the four-query rankings significantly improve the training speed and returns in all environments, especially the Two-Switch variant. In this scenario, the four-query rankings of 80\% correctness dramatically raise the training returns to the optimal. The policy with rewards from single-query rankings of 70\% correctness fails, while the four-query rankings of 70\% accuracy considerably improve the individual reward quality and train some useful policies.

\subsubsection{\textbf{\textit{LLM Two-Query Evaluation}}}

As discussed above, the q3 version of the Llama3.1-70B is faster and more accessible than the full-sized version but generates more errors and has a flawed performance when training credit-aware individual rewards using a single query per state pair. This section shows that the learning speed can be accelerated with less variance and the training return can be raised to the optimal if using one more query to rank each state pair, as demonstrated by the learning curves averaged over 3 random seeds in Fig. ~\ref{fig:2query}. In the Pistonball environment, since the policy trained with single-query Llama3.1-70B:q3 rankings is already with the fastest learning speed, least variance and optimal training returns, the improvement from the multi-query approach is limited and the two-query variation remains on par with it.

\section{Conclusions}
This work leverages LLMs to handle the critical challenge of credit assignment in MARL in environments with sparse rewards. This LCA method decomposes sparse team rewards into dense, agent-specific ones by using LLM to evaluate each agent's actions in the contexts of collaboration. The potential-based reward-shaping mechanism mitigates the impact of LLM hallucination, enhancing the robustness and reliability of our method. Our extensive experiments demonstrate that multi-agent collaboration policies trained with our LLM-guided individual rewards achieve faster convergence and higher policy returns compared to state-of-the-art MARL baselines. Experiments also show the resilience of LCA to ranking errors. Therefore, without significant performance degradation, LCA is applicable to smaller and more accessible language models.

\section*{Acknowledgement}
We would like to acknowledge the support from Honda under grant 58629.1.1012949, from DARPA under ANSR grant FA8750-23-2-1015 and Prime Contract No. HR00112490409, as well as from ONR under CAI grant N00014-23-1-2840.

\nocite{langley00}

\bibliography{example_paper}
\bibliographystyle{conference2025}

\newpage
\appendix
\onecolumn
\section{Individual-Reward Rollout over an Episode}
\label{sec:rollout}
\begin{figure}[ht]
\vskip 0.2in
\begin{center}
\centerline{\includegraphics[width=1\linewidth]{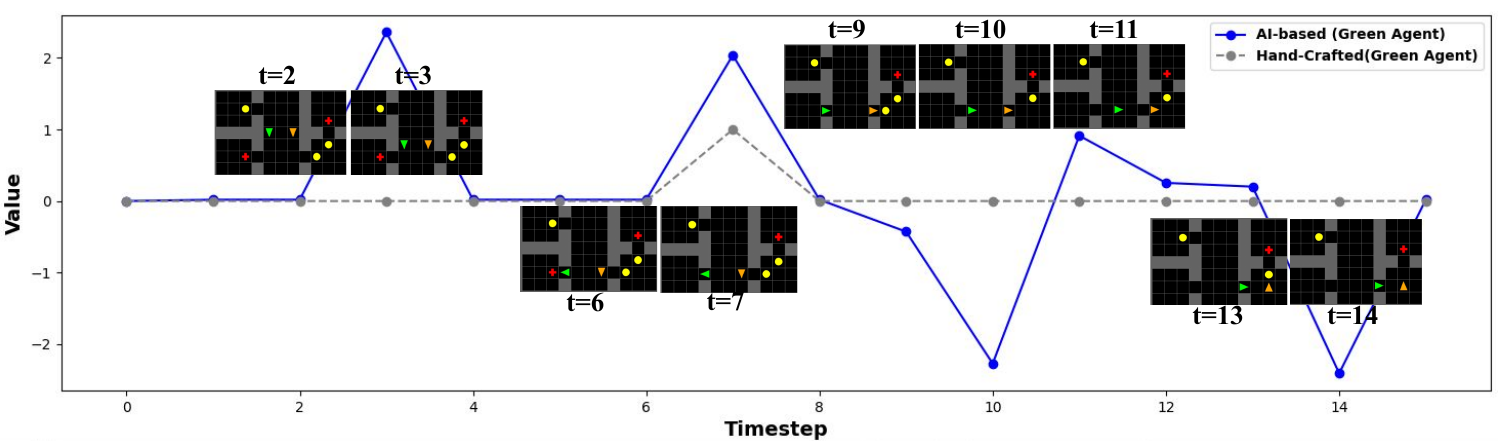}}
\caption{Rolling out individual rewards (blue line) over states of an episode from time steps 0 to 15 in the Victim-Rubble environment. The individual rewards here are the potential-based rewards trained with single-query GPT4o rankings.}
\label{fig:rollout}
\end{center}
\vskip -0.2in
\end{figure}

We plotted the green agent's individual rewards at states from a continuous episode in the Victim-Rubble environment. The individual rewards here are trained with single-query GPT4o rankings. Compared with the default sparse team reward (grey line), we can see that LCA successfully generates dense individual rewards evaluating individual actions in the contexts of collaboration. Besides giving positive rewards when the green agent makes significant progress (ie. saving victims) like simple hand-crafted reward functions do, LCA also rewards the green agent when it makes a critical turn or movement to the correct target (t=3, 11 in Fig. \ref{fig:rollout}). Meanwhile, LCA individual rewards punish the green agent not only when it takes the wrong action, but also when its teammate makes significant progress but it does nothing special (t=10, 14). It seems that the LLM tends to push the green agent to make progress, effectively avoiding lazy agents.

\end{document}